%% file: paper_submit.tex
\documentclass[runningheads]{llncs}
\usepackage[T1]{fontenc}
\usepackage{bm} 
\usepackage{graphicx}
\usepackage{mathrsfs}
\usepackage{array}
\usepackage{amsfonts, amssymb}
\usepackage{amsmath}
\usepackage[mathscr]{eucal}
\usepackage{mathtools}
\usepackage{diagbox}
\usepackage{float}
\usepackage{bbding}
\usepackage{xcolor}
\usepackage{hyperref}
\usepackage{ulem}
\usepackage{makecell}
\usepackage{dsfont}
\usepackage{multirow}
\usepackage{cite}
\usepackage{ulem}
\usepackage{booktabs}
\usepackage{wrapfig}

\newcommand{\norm}[1]{\left\lVert#1\right\rVert}

\newcommand{\mmid}{\!\mid\!}

\usepackage{CJKutf8}
\usepackage{changes}

\usepackage{algorithm}
\usepackage{algpseudocode}
\usepackage{stfloats}
\usepackage{tabularx}
\usepackage{colortbl}
\usepackage{pifont}

\makeatletter

\newcommand{\Rmnum}[1]{\expandafter\@slowromancap\romannumeral #1@}
\makeatother

\begin{document}
\title{Open World MRI Reconstruction with Bias-Calibrated Adaptation}

\titlerunning{Open World MRI Reconstruction with Bias-Calibrated Adaptation}
\author{Jiyao Liu\inst{1} \and
Shangqi Gao \inst{2}$^{\text{\Envelope}}$  \and 
Lihao Liu\inst{3}  \and 
Junzhi Ning \inst{3}  \and 
Jinjie Wei\inst{1}  \and \\
Junjun He\inst{3}  \and 
Xiahai Zhuang\inst{1}$^{\text{\Envelope}}$ \and
Ningsheng Xu\inst{1}
}

\authorrunning{J Liu et al. }

\institute{Fudan University, Shanghai, China \and
University of Cambridge, Cambridge, United Kingdom \and
Shanghai Artificial Intelligence Laboratory, Shanghai, China}

\maketitle              

\input{section/0_abstract.tex}

\input{section/1_intro}

\input{section/2_methodology}

\input{section/3_experiments}

\input{section/4_conclusion}

\clearpage
\newpage

\bibliographystyle{splncs04}
\bibliography{bib}  

\end{document}

%% file: section/0_abstract.tex
\begin{abstract}
Real-world MRI reconstruction systems face the open-world challenge: test data from unseen imaging centers, anatomical structures, or acquisition protocols can differ drastically from training data, causing severe performance degradation.
Existing methods struggle with this challenge. To address this, we propose \textbf{BiasRecon}, a bias-calibrated adaptation framework grounded in the \textit{minimal intervention principle: preserve what transfers, calibrate what does not.} Concretely, BiasRecon formulates open-world adaptation as an alternating optimization framework that jointly optimizes three components: (1) \textit{frequency-guided prior calibration} that introduces layer-wise calibration variables to selectively modulate frequency-specific features of the pre-trained score network via self-supervised k-space signals, (2) \textit{score-based denoising} that leverages the calibrated generative prior for high-fidelity image reconstruction, and (3) \textit{adaptive regularization} that employs Stein's Unbiased Risk Estimator to dynamically balance the prior-measurement trade-off, matching test-time noise characteristics without requiring ground truth.
By intervening minimally and precisely through this alternating scheme, BiasRecon achieves robust adaptation with fewer than 100 tunable parameters. Extensive experiments across four datasets demonstrate state-of-the-art performance on open-world reconstruction tasks.

\keywords{Open-World Learning \and MRI Reconstruction \and Test-Time Adaptation}
\end{abstract}

%% file: section/1_intro.tex
\section{Introduction}
\label{sec:intro}

\input{section/fig_tex/distribution_shift}

MRI reconstruction aims to recover high-quality images from undersampled k-space measurements. Traditional approaches rely on hand-crafted priors such as wavelets \cite{qu2012undersampled}, total variation \cite{block2007undersampled}, and low-rank constraints \cite{yang2016sparse}. Deep learning has significantly advanced the field through supervised methods \cite{aggarwal2018modl, yang2018admm}, unsupervised approaches \cite{ulyanov2018deep, yaman2021zero}, and generative prior-based methods \cite{liu2020highly, wang2021denoising, chung2022score, chungdecomposed}.

However, clinical deployment presents a fundamentally different challenge. In practice, reconstruction models must operate under \textbf{open-world conditions}, where the test data distribution is unknown and can differ drastically from the training distribution. Such distribution shifts may arise from variations in anatomical structures, imaging protocols, scanner characteristics, or their combinations, as Fig.~\ref{distribution_shift1}(a) illustrates. In clinical practice, such open-world scenarios are the norm rather than the exception.

Existing reconstruction methods experience significant performance degradation under such distribution shifts, as Fig.~\ref{fig:comparison} demonstrates. We hypothesize that this degradation arises from two sources: (1) the learned image prior becomes misaligned with the target distribution, and (2) hyperparameters tuned on training data become suboptimal for new domains.

A natural question arises: how can we adapt a pre-trained model to unknown distributions with only single-image supervision? We argue that the key insight lies in recognizing that \textbf{\textit{not all learned knowledge degrades equally under distribution shifts}}. Through transferability analysis (Sec.~\ref{PA}), we find that deeper layers and low-frequency components are domain-specific, while shallow layers and high-frequency components remain domain-agnostic, as visualized in Fig.~\ref{distribution_shift1}(b,c). Therefore, efficient adaptation must jointly address \textit{what} to adapt and \textit{how much} to adapt.

Building on this insight, we propose \textbf{BiasRecon} (Fig.~\ref{framework}), a bias-calibrated adaptation framework grounded in the \textit{minimal intervention principle}: preserve transferable knowledge while calibrating only the components that cause degradation. BiasRecon operationalizes this principle through an alternating optimization framework that directly addresses both hypothesized sources of degradation. To correct the misaligned prior, \textit{frequency-guided prior calibration} adaptively modulates layer-wise and frequency-specific components, achieving effective adaptation with minimal parameters, while \textit{score-based denoising} leverages the calibrated prior for reconstruction. To address the suboptimal hyperparameters, \textit{adaptive regularization} dynamically adjusts the prior-measurement trade-off to match the statistical characteristics of each test sample, using a self-supervised criterion that requires no ground truth. This principled design enables BiasRecon to achieve robust open-world adaptation with fewer than 100 tunable parameters.

\textbf{Contributions.} First, we identify through transferability analysis that deeper layers and low-frequency components are domain-specific while shallow layers and high-frequency components remain transferable, motivating the \textit{minimal intervention principle} for open-world adaptation. Second, we propose BiasRecon, an alternating optimization framework with fewer than 100 tunable parameters that operationalizes this principle through frequency-guided prior calibration and adaptive regularization. Third, we demonstrate state-of-the-art performance on open-world MRI reconstruction.

%% file: section/fig_tex/distribution_shift.tex
\begin{figure*}[t!]
  \centering
  \includegraphics[width=0.85\textwidth]{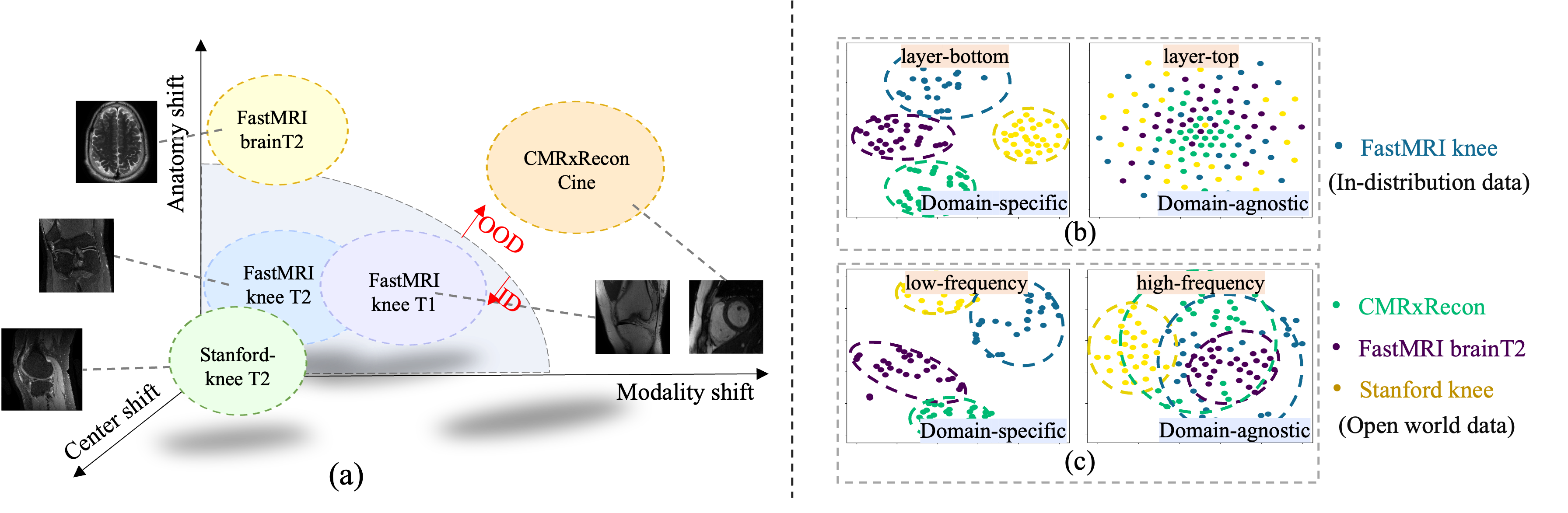}
  \caption{\textbf{Transferability analysis.} (a) Open-world distribution shifts from anatomy, center, and modality changes. Models trained on FastMRI knee (inside dashed box) are tested on three OOD scenarios. (b) t-SNE visualization of features from bottom (deeper) and top (shallow) layers across different domains. (c) Frequency decomposition analysis of low-frequency and high-frequency components across different domains.}
  \label{distribution_shift1}
  \vspace{-0.5em}
\end{figure*}

%% file: section/2_methodology.tex
\section{Method}
\label{sec:method}

\input{section/fig_tex/framework}

\subsection{Bias-Calibrated Adaptation Framework}
\label{sec:framework}

\textbf{MRI Forward Model.} In parallel accelerated MRI, the observation obtained in multi-coil k-space $\mathbf{y}_c \in \mathbb{C}^{N_y}$ is mathematically expressed as $\mathbf{y}_c = \mathbf{A}_c \mathbf{x} + \mathbf{n}$, where $\mathbf{x} \in \mathbb{C}^{N_x}$ is the reconstructed image, $\mathbf{n}$ models Gaussian noise, and $\mathbf{A}_c = \boldsymbol{\mathcal{M}} \boldsymbol{\mathcal{F}} \mathbf{S}_c$ is the forward operator for the $c$-th coil. Here, $\mathbf{S}_c$ encodes the coil sensitivity map, $\boldsymbol{\mathcal{F}}$ denotes the discrete Fourier transform, and $\boldsymbol{\mathcal{M}}$ is the undersampling mask \cite{geethanath2013compressed}. For a multi-coil system with $C$ coils, we write $\mathbf{y} = \mathbf{A} \mathbf{x} + \mathbf{n}$.

\textbf{Bias-Calibrated Adaptation.} We adopt score-based generative models \cite{chung2022score} as our prior estimator, where a neural network $\boldsymbol s_{\theta^*}(\mathbf{x})$ approximates the score function $\nabla_{\mathbf{x}} \log p(\mathbf{x})$. To tackle open-world MRI reconstruction, we introduce a calibration variable $\boldsymbol\delta$ that modulates the pre-trained score network to accommodate distribution shifts, yielding $\boldsymbol s_{\theta^*}(\mathbf{x}; \boldsymbol{\delta}) \approx \nabla_{\mathbf{x}} \log p(\mathbf{x}\mmid\boldsymbol{\delta})$.
We achieve reconstruction by jointly optimizing $\mathbf{x}$ and $\boldsymbol{\delta}$ given observation $\mathbf{y}$:
\begin{equation}
  (\mathbf x, \boldsymbol{\delta}) \in \arg\max_{\mathbf x, \boldsymbol{\delta}}p(\mathbf x,\boldsymbol{\delta}\mmid\mathbf y) \in \arg\max_{\mathbf x, \boldsymbol{\delta}}p(\mathbf{y}\mmid\mathbf x)p(\mathbf x\mmid\boldsymbol{\delta})p(\boldsymbol{\delta}),
  \label{MAP}
\end{equation}
where $p(\mathbf y\mmid\mathbf x)=\mathcal N(\mathbf y\mmid\mathbf A\mathbf x, \gamma^{-1}\mathbf{I})$ and $p(\mathbf{x}\mmid\boldsymbol{\delta})$ is determined through the score function. Taking the negative log and introducing a proximal term, we split the problem into three alternating subproblems:
\begin{equation}
    \label{equ:7}
    \small
    \begin{cases}
    P1: \boldsymbol{\delta}^{t-1} \in  \arg\max_{\boldsymbol\delta} \underbrace{\log p(\mathbf x^t\mmid\boldsymbol\delta) + \log p(\boldsymbol{\delta})}_{Prior\ calibration}, \\
    P2: \dot{\mathbf x}^{t-1} \in \arg\min_{\mathbf x^t} \underbrace{\frac{1}{2}\|\dot{\mathbf{x}}^t - \mathbf{x}^t \|_2^2 - \log p(\mathbf x^t\mmid \boldsymbol\delta^{t-1})}_{Denoising},\\
    P3: \hat{\mathbf x}^{t-1} = \arg\min_{\hat{\mathbf x}}\frac{\gamma}{2} \underbrace{\|\mathbf{y}-\mathbf A\hat{\mathbf x}\|_2^2}_{Data\ Fidelity} + \frac{1}{2} \underbrace{\|\dot{\mathbf x}^{t-1} - \hat{\mathbf x}\|_2^2}_{Proximity}.
    \end{cases}
\end{equation}

For P2 in Eq.~\eqref{equ:7}, we derive a one-step update:
\begin{equation}
\dot{\mathbf x}^{t-1} = \mathbf{x}^{t} + (\sigma^t)^2 \nabla_{\mathbf x}\log p(\mathbf{x}\mmid\boldsymbol{\delta}^{t-1})|_{\mathbf{x}=\mathbf{x}^t},
\label{x_itera}
\end{equation}
where the score function is estimated by a bias-calibrated network $\boldsymbol s_{\theta^*}(\mathbf{x}^t,t; \boldsymbol{\delta}^{t-1})$ and $(\sigma^t)^2$ is derived from Tweedie's formula \cite{knoll2019assessment}. For P3 in Eq.~\eqref{equ:7}, we use conjugate gradient \cite{chungdecomposed} to efficiently solve the data fidelity term.

\subsection{Frequency-guided Prior Calibration}
\label{PA}

\begin{figure}[t]
    \centering
    \begin{minipage}[t]{0.45\linewidth}
        \centering
        \includegraphics[width=\linewidth]{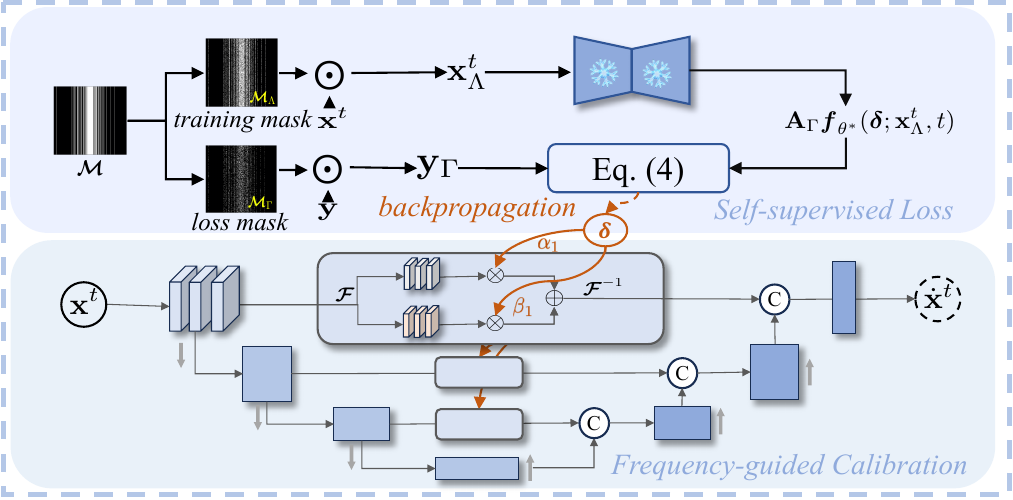}
        \caption{Illustration of frequency-guided prior calibration.}
        \label{fig:calibration}
    \end{minipage}
    \hfill
    \begin{minipage}[t]{0.52\linewidth}
        \centering
        \includegraphics[width=\linewidth]{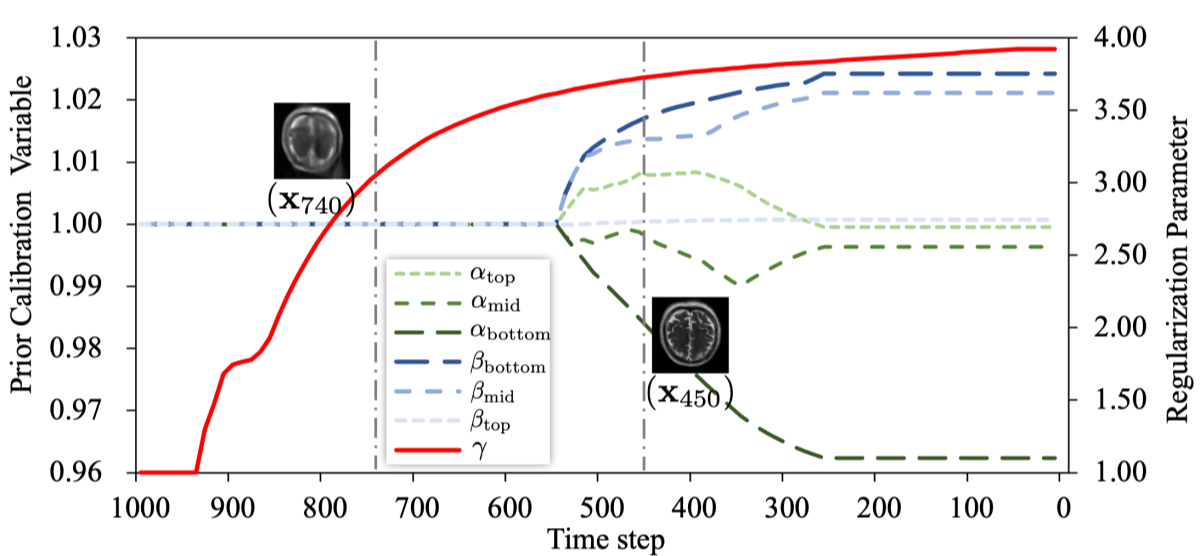}
        \caption{Dynamic evolution of calibration variables and regularization parameter during the reconstruction process.}
        \label{param}
    \end{minipage}
\end{figure}

\textbf{Transferability Analysis.} To understand how distribution shifts affect the denoising network in P2 of Eq.~\eqref{equ:7}, we analyze feature representations through t-SNE visualization, as shown in Fig.~\ref{distribution_shift1}(b,c). We find that \textit{deeper layers and low-frequency components exhibit domain-specific characteristics (encoding anatomical structures like knee vs. brain), while shallow layers and high-frequency components remain domain-agnostic (capturing transferable edges and textures)}. This observation motivates us to design a layer-wise and frequency-specific modulation strategy that selectively calibrates low-frequency components in deeper layers while preserving high-frequency representations in shallow layers.

\textbf{Parameter-efficient Calibration Strategy.} Motivated by the above analysis, we instantiate the calibration variable $\boldsymbol{\delta}$ as a compact parameter vector for the U-Net architecture \cite{ronneberger2015u} of our score-based network. Specifically, for each skip connection layer $l$, we decompose features via Fourier transformation and introduce two learnable scalars $\alpha_l, \beta_l \in [0,2]$ to modulate low- and high-frequency components, respectively. The calibration variable is thus defined as $\boldsymbol{\delta} = [\alpha_1, \beta_1, \ldots, \alpha_L, \beta_L]^\top \in \mathbb{R}^{2L}$, introducing just $2L$ parameters for an $L$-layer network, enabling efficient frequency-aware adaptation, as illustrated in Fig.~\ref{fig:calibration}.

\textbf{Self-supervised Optimization.} With $\boldsymbol{\delta}$ defined above, for P1 in Eq.~\eqref{equ:7}, we optimize $\boldsymbol{\delta}$ without ground truth by partitioning the measurement mask into two complementary Gaussian 2D subsampling masks ($\boldsymbol{\mathcal M}=\boldsymbol{\mathcal M}_\Lambda + \boldsymbol{\mathcal M}_\Gamma$). Through k-space projection and probabilistic modeling, we obtain:
\begin{equation}
\mathcal{L}_{\mathrm{SSL}}(\boldsymbol{\delta}) = \frac{\tau}{2}\norm{\mathbf{y}_{\Gamma} - \mathbf{A}_{\Gamma} \boldsymbol{f}_{\theta^*} \left(\boldsymbol{\delta};\mathbf{x}_\Lambda^t, t\right)}_2^2,
\label{equ:delta_update}
\end{equation}
where $\boldsymbol{f}_{\theta^*}$ represents the reconstruction from P2 and P3, and $\tau$ controls precision. Therefore, P1 in Eq.~\eqref{equ:7} can be reformulated as:
\begin{equation}
    \boldsymbol{\delta}^{t-1} \in \arg\min_{\boldsymbol\delta}  \mathcal{L}_{\mathrm{SSL}}(\boldsymbol{\delta}) - \log p(\boldsymbol{\delta}),
    \label{delta_final}
\end{equation}
where $p(\boldsymbol{\delta}) = \mathcal{N}(1, 1)$ ensures $\boldsymbol{\delta} = 1$ preserves the original network.

\subsection{Regularization Parameter Adaptation}
\label{CA}

For P3 in Eq.~\eqref{equ:7}, the regularization parameter $\gamma$ balances measurement consistency and prior constraints. Manual selection is suboptimal for open-world scenarios since $\gamma$ relates to the noise level, which should be image-specific and updated adaptively. We employ Stein's Unbiased Risk Estimator (SURE) \cite{zhussip2019extending} with Monte Carlo approximation \cite{ramani2008monte} to estimate MSE without ground truth:
\begin{align}
    \label{equ:SURE}
    \mathcal L_\mathrm{Reg}(\gamma) = \frac{\|\mathbf x_{\mathrm{zf}} - \boldsymbol h(\gamma;\mathbf x^t)\|^2}{N\epsilon}\boldsymbol\mu^T(\boldsymbol h(\gamma;\mathbf x^t+\epsilon\boldsymbol\mu)-\boldsymbol h(\gamma;\mathbf x^t)),
\end{align}
where $\boldsymbol\mu \sim \mathcal{N}(0, \mathbf{I})$, $\epsilon$ is the perturbation scale, and $\boldsymbol h(\gamma;\cdot)$ represents the compositional mapping of P2 and P3 from Eq.~\eqref{equ:7}. To prevent overfitting, we employ early stopping based on sliding window convergence:
\begin{equation}
    \label{convergence_criterion}
    \mathcal E(\mathcal L_\mathrm{Reg}) = 1 - \frac{\sum_{t'=t-k+1}^{t}\mathcal L^{t'}_\mathrm{Reg}}{\sum_{t'=t-2k+1}^{t-k}\mathcal L^{t'}_\mathrm{Reg}},
\end{equation}
where updating stops when $\mathcal E(\mathcal L_\mathrm{Reg}) < \tau_\mathrm{Reg}$.

%% file: section/fig_tex/framework.tex
\begin{figure*}[t!]
  \centering 
  \includegraphics[width=\textwidth]{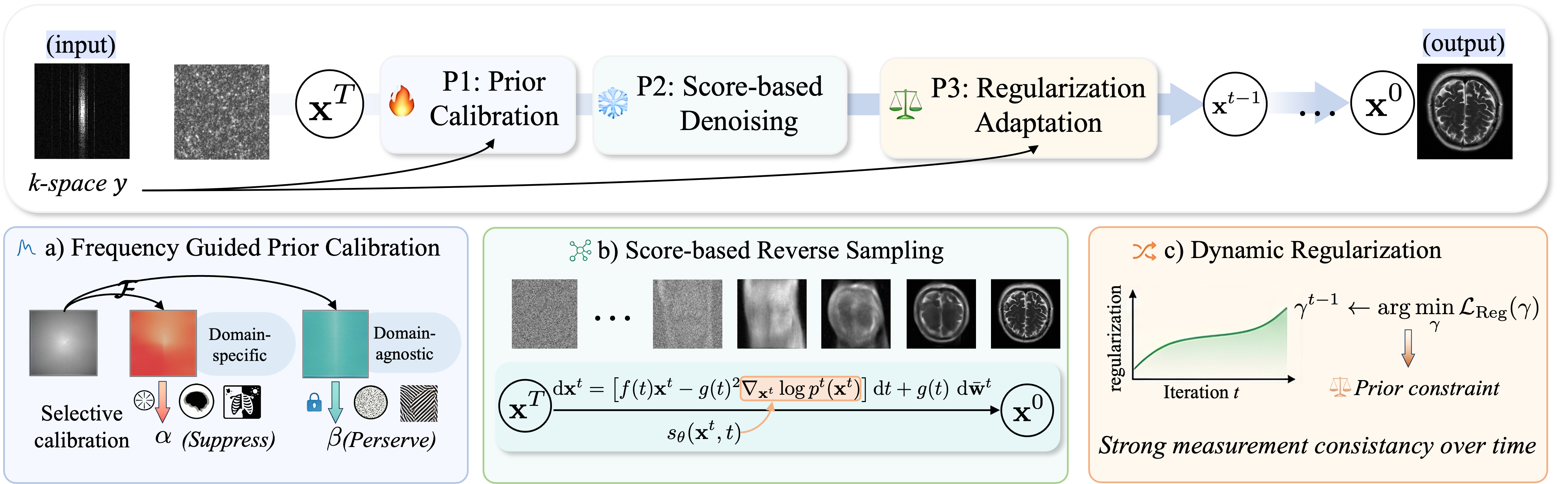}
  \caption{\textbf{Overview of BiasRecon framework.} The top panel illustrates the alternating optimization process of BiasRecon.
  The framework consists of three key components: a) Frequency-guided Prior Calibration, which adaptively adjusts feature distributions through parameter-efficient modulation; b) Score-based reverse sampling process as the foundation; and c) Regularization Parameter Adaptation, which dynamically balances measurement consistency and prior constraints.}
  \vspace{-1em}
  \label{framework}
  \end{figure*} 

%% file: section/3_experiments.tex
\section{Experiments}
\label{sec:experiment}

\subsection{Experimental Setup}

\textbf{Datasets.} We evaluate on four multi-coil MRI datasets. \textit{FastMRI knee} \cite{zbontar2018fastmri} serves as the training set (973 volumes), discarding the first and last 5 slices of each volume with small objects. We directly evaluate on three open-world test sets without retraining: \textit{FastMRI-Brain} (Cross-Anatomy), \textit{Stanford Knee} (Cross-Center) \cite{ong2018mridata}, and \textit{CMRxRecon} (Combined shift) \cite{wang2024cmrxrecon}. For each test set, 100 volumes are randomly selected with 3 slices per volume. Input images are unified to $320 \times 320$ through cropping and padding. Coil sensitivity maps are estimated by ESPiRiT \cite{uecker2014espirit}.

\textbf{Baselines and Evaluation.} We compare against TV \cite{block2007undersampled}, ConvDecoder \cite{darestani2021accelerated}, MoDL \cite{aggarwal2018modl}, and DDS \cite{chungdecomposed}. Reconstruction quality is evaluated by PSNR, SSIM, and LPIPS \cite{zhang2018perceptual}.

\textbf{Training and Inference.} We train a score-based model on FastMRI knee for 1M iterations with batch size 4 and learning rate 1e-4. Complex data is processed using MVUE \cite{jalal2021robust}.
We test Gaussian1D and Uniform1D undersampling at 4× and 8× acceleration. During inference, calibration variable $\boldsymbol{\delta}$ and regularization parameter $\gamma$ are updated via Adam optimizer. We employ DDIM sampling \cite{song2020denoising} with 100 timesteps to accelerate the reverse process. $\boldsymbol{\delta}$ is initialized to identity and $\gamma$ to 1. We set $\tau_\mathrm{Reg} = 0.001$ with window size $k=5$.

\subsection{Open-World Evaluation}
    \begin{table*}[ht]
        \vspace{-1em}
        \scriptsize
        \setlength{\tabcolsep}{3.5pt}
        \renewcommand{\arraystretch}{1.0}
        \caption{Open-world MRI reconstruction performance under 8$\times$ acceleration with Gaussian 1D sampling.}
        \label{tab:Comparision}
        \centering
        \resizebox{\textwidth}{!}{
        \begin{tabular}{cccccccc}
            \toprule
            \multirow{2}{*}{Method}  &\multicolumn{3}{c}{FastMRI Knee (In-Distribution)} &\multicolumn{3}{c}{FastMRI-Brain (Cross-Anatomy)} \\
            \cmidrule{2-7}
            & PSNR$\uparrow$  & SSIM$\uparrow$ &LPIPS$\downarrow$ & PSNR$\uparrow$  & SSIM$\uparrow$ &LPIPS$\downarrow$\\
            \midrule
            Zero-filling& 19.83±1.52  & 0.691±0.025 & 0.329±0.031 & 23.72±1.83 & 0.785±0.028 & 0.236±0.022\\
            \hline\rowcolor{gray!30} \multicolumn{7}{c}{\textit{Conventional CS method}}\\
            TV & 26.04±1.63 & 0.771±0.023 & 0.313±0.025 & 28.14±1.86 & 0.817±0.027 & 0.187±0.024\\
            \hline\rowcolor{gray!30} \multicolumn{7}{c}{\textit{Internal deep image prior}}\\
            ConvDecoder & 29.97±2.15 & 0.745±0.029 & 0.257±0.027 & 31.27±2.73 & 0.876±0.022 & 0.141±0.022\\
            \hline\rowcolor{gray!30} \multicolumn{7}{c}{\textit{Supervised Method}}\\
            MoDL & 31.67±1.98 & 0.773±0.028 & 0.215±0.024 & 32.53±0.99 & 0.807±0.064 & 0.137±0.019\\
            \hline\rowcolor{gray!30} \multicolumn{7}{c}{\textit{External deep image prior}}\\
            DDS & 31.64±1.89 & 0.780±0.031 & 0.200±0.021 & 32.46±1.63 & 0.864±0.029 & 0.140±0.016\\
            Ours & \textbf{31.76±2.01} & \textbf{0.792±0.032} & \textbf{0.195±0.018} & \textbf{33.65±1.42} & \textbf{0.885±0.019} & \textbf{0.136±0.025}\\
            \bottomrule
        \end{tabular}}

        \vspace{0.2em}
        \resizebox{\textwidth}{!}{
        \begin{tabular}{cccccccc}
            \toprule
            \multirow{2}{*}{Method}  &\multicolumn{3}{c}{Stanford Knee (Cross-Center)} &\multicolumn{3}{c}{CMRxRecon (Combined)} \\
            \cmidrule{2-7}
            & PSNR$\uparrow$  & SSIM$\uparrow$ &LPIPS$\downarrow$ & PSNR$\uparrow$  & SSIM$\uparrow$ &LPIPS$\downarrow$\\
            \midrule
            Zero-filling& 19.51±1.48  & 0.691±0.027 & 0.327±0.029 & 22.82±1.76 & 0.676±0.031 & 0.164±0.021\\
            \hline\rowcolor{gray!30} \multicolumn{7}{c}{\textit{Conventional CS method}}\\
            TV & 24.12±1.72 & 0.788±0.026 & 0.306±0.028 & 31.15±1.92 & 0.804±0.029 & 0.182±0.023\\
            \hline\rowcolor{gray!30} \multicolumn{7}{c}{\textit{Internal deep image prior}}\\
            ConvDecoder & 30.84±2.25 & 0.753±0.031 & 0.264±0.028 & 34.71±2.42 & 0.854±0.025 & 0.133±0.021\\
            \hline\rowcolor{gray!30} \multicolumn{7}{c}{\textit{Supervised Method}}\\
            MoDL & 30.66±1.85 & 0.803±0.029 & 0.203±0.023 & 32.98±1.94 & \textbf{0.877±0.021} & 0.143±0.022\\
            \hline\rowcolor{gray!30} \multicolumn{7}{c}{\textit{External deep image prior}}\\
            DDS & 29.23±1.78 & 0.822±0.027 & 0.198±0.022 & 33.72±1.96 & 0.865±0.026 & 0.138±0.019\\
            Ours & \textbf{31.29±1.92} & \textbf{0.846±0.028} & \textbf{0.185±0.021} & \textbf{35.32±2.12} & \textbf{0.878±0.024} & \textbf{0.116±0.020}\\
            \bottomrule
        \end{tabular}}
        \label{table:comparision1}
        \vspace{-3.5em}
    \end{table*}

\begin{figure*}
    \centering
    \includegraphics[width=0.97\textwidth]{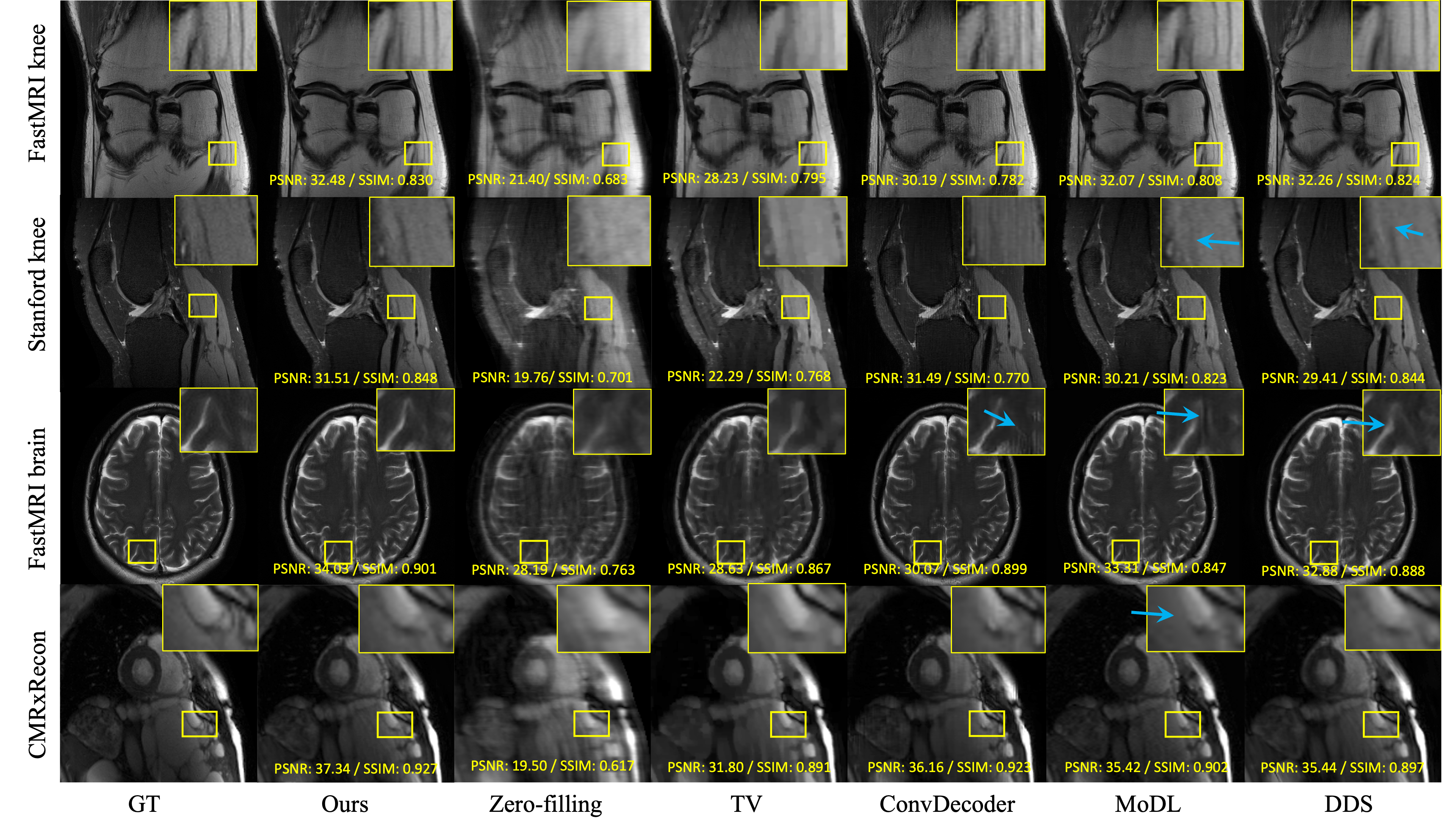}
    \caption{Qualitative comparison of MRI reconstruction results under 8$\times$ Gaussian 1D undersampling.}
    \label{fig:comparison}
    \vspace{-2em}
\end{figure*}

\subsubsection{Robustness against Distribution Shifts.}

Table~\ref{table:comparision1} demonstrates BiasRecon's effectiveness across diverse distribution shifts. On the in-distribution FastMRI knee dataset, our method slightly outperforms baselines by adapting to image-specific noise levels. More importantly, BiasRecon achieves the best performance across all open-world scenarios: +1.19 dB PSNR over DDS on \textit{Cross-Anatomy}, +2.06 dB on \textit{Cross-Center}, and +1.60 dB on \textit{Combined}, which is the most challenging scenario involving cross-anatomy, cross-center, and cross-modality shifts simultaneously. These consistent improvements validate our two hypothesized sources of degradation: frequency-guided prior calibration corrects the misaligned prior, while adaptive regularization adjusts the prior-measurement trade-off for unseen acquisition conditions. Qualitative results in Fig.~\ref{fig:comparison} further confirm that BiasRecon produces reconstructions most consistent with ground truth.

    \begin{table*}[ht]
        \tiny
        \setlength{\tabcolsep}{3.5pt}
        \renewcommand{\arraystretch}{1.0}
        \caption{Robustness against different sampling patterns and acceleration factors on open-world FastMRI-brain dataset. Comparison includes both Gaussian 1D and Uniform 1D sampling at 4$\times$ and 8$\times$ acceleration rates.}
        \label{table:robust_operation}
        \centering
        \resizebox{\textwidth}{!}{
        \begin{tabular}{llcccc}
        \toprule
        \multirow{2}{*}{Method} & \multirow{2}{*}{Metric} & \multicolumn{2}{c}{Gaussian 1D} & \multicolumn{2}{c}{Uniform 1D} \\
        \cmidrule(lr){3-4} \cmidrule(lr){5-6}
        & & 4$\times$ & 8$\times$ & 4$\times$ & 8$\times$ \\
        \midrule
        \multirow{3}{*}{TV} & PSNR & 33.65 ± 1.29 & 28.14 ± 1.86 & 28.59 ± 1.69 & 25.71 ± 1.61 \\
        & SSIM & 0.868 ± 0.017 & 0.817 ± 0.027 & 0.801 ± 0.031 & 0.725 ± 0.043 \\
        & LPIPS & 0.136 ± 0.017 & 0.187 ± 0.024 & 0.174 ± 0.028 & 0.254 ± 0.043 \\
        \hline
        \multirow{3}{*}{ConvDecoder} & PSNR & 33.19 ± 2.41 & 31.27 ± 2.73 & 31.94 ± 2.35 & 26.27 ± 2.86 \\
        & SSIM & \textbf{0.892 ± 0.020} & \underline{0.876 ± 0.022} & \underline{0.860 ± 0.034} & 0.728 ± 0.070 \\
        & LPIPS & 0.116 ± 0.026 & 0.141 ± 0.022 & 0.135 ± 0.031 & 0.247 ± 0.052 \\
        \hline
        \multirow{3}{*}{MoDL} & PSNR & 33.59 ± 1.637 & 32.53 ± 0.99 & \underline{31.96 ± 2.056} & 27.08 ± 1.919 \\
        & SSIM & 0.871 ± 0.022 & 0.807 ± 0.064 & 0.797 ± 0.064 & 0.659 ± 0.077 \\
        & LPIPS & 0.120 ± 0.032 & 0.137± 0.019 & 0.125 ± 0.021 & \textbf{0.020 ± 0.040} \\
        \hline
        \multirow{3}{*}{DDS} & PSNR & \underline{34.10 ± 1.97} & \underline{32.56 ± 1.63} & 30.99 ± 1.79 & \underline{27.23 ± 2.03} \\
        & SSIM & 0.883 ± 0.028 & 0.864 ± 0.029 & 0.851 ± 0.031 & \underline{0.736 ± 0.049} \\
        & LPIPS & 0.114 ± 0.015 & 0.140 ± 0.016 & 0.135 ± 0.020 & 0.223 ± 0.040 \\
        \hline
        \multirow{3}{*}{Ours} & PSNR & \textbf{35.03 ± 2.01} & \textbf{33.65±1.42} & \textbf{32.72 ± 1.92} & \textbf{28.40 ± 2.22} \\
        & SSIM & \underline{0.889 ± 0.032} & \textbf{0.885±0.019} & \textbf{0.869 ± 0.033} & \textbf{0.743 ± 0.041} \\
        & LPIPS & \textbf{0.110 ± 0.018} & \textbf{0.136±0.025} & \textbf{0.134 ± 0.023} & \underline{0.211 ± 0.046} \\
        \bottomrule
        \end{tabular}}
    \end{table*}

\subsubsection{Robustness against undersampling.}
We evaluated the robustness against different undersampling operators, including Uniform1D and Gaussian2D, with acceleration factors of \( 4 \times \) and \( 8 \times \). For experimental fairness, the MoDL model was trained on the FastMRI knee dataset using different sampling operators at these acceleration factors (\( 4 \times \) and \( 8 \times \)). Table \ref{table:robust_operation} presents the quantitative results. Our method outperforms others in 10 out of 12 studies and achieves the second best performance in the rest 2 studies, demonstrating greater robustness across various sampling patterns and acceleration factors.

\subsection{Parameter Analysis}

Fig.~\ref{param} shows that while $\boldsymbol{\alpha}$ and $\boldsymbol{\beta}$ in top layers remain stable, bottom layers exhibit pronounced changes: $\boldsymbol{\alpha}$ decreases while $\boldsymbol{\beta}$ increases, indicating suppression of domain-specific low-frequency components while preserving high-frequency details. The gradual increase in $\gamma$ reflects the transition from prior-guided denoising to measurement-consistent refinement as reconstruction progresses. 

\subsection{Ablation Study}

\begin{table}[t]
    \caption{Ablation study on Cross-Center shift under 4$\times$ Gaussian 1D acceleration. FPC: Frequency-guided Prior Calibration; RPA: Regularization Parameter Adaptation.}
    \label{tab:ablation_2factor_StanfordKnee}
    \centering
    \scriptsize
    \setlength{\tabcolsep}{3pt}
    \renewcommand{\arraystretch}{0.95}
    \resizebox{\columnwidth}{!}{
    \begin{tabular}{lccc}
        \toprule
        Method & PSNR$\uparrow$  & SSIM$\uparrow$ & LPIPS$\downarrow$ \\
        \midrule
        Baseline & $30.23\pm3.57$ & $0.844\pm0.035$ & $0.187\pm0.028$ \\
        w/o RPA & $31.28\pm3.22(\uparrow1.05)$ & $0.855\pm0.035(\uparrow.011)$ & $0.183\pm0.029(\downarrow.004)$ \\
        w/o FPC & $32.51\pm4.04(\uparrow2.28)$ & $0.853\pm0.058(\uparrow.009)$ & $0.157\pm0.065(\downarrow.030)$ \\
        \textbf{Ours (FPC + RPA)} & $\mathbf{33.17\pm3.11}(\uparrow2.94)$ & $\mathbf{0.866\pm0.041}(\uparrow.022)$ & $\mathbf{0.138\pm0.027}(\downarrow.049)$ \\
        \bottomrule
    \end{tabular}}
    \vspace{-1.5em}
\end{table}

Table~\ref{tab:ablation_2factor_StanfordKnee} validates the effectiveness of our two adaptation components under cross-center distribution shift. Both components contribute meaningfully to the reconstruction quality, demonstrating their complementary nature. \textit{Frequency-guided prior calibration} (+1.05 dB) adapts the learned prior by calibrating low-frequency components to accommodate variations in contrast, intensity distributions, and scanner-specific artifacts across different imaging centers. \textit{Regularization parameter adaptation} (+2.28 dB) dynamically adjusts the prior-measurement trade-off to match test-time noise characteristics and k-space sampling quality. The full BiasRecon framework achieves +2.94 dB improvement, demonstrating that both components work synergistically.

%% file: section/4_conclusion.tex
\section{Conclusion}
\label{sec:conclusion}

We present BiasRecon, a bias-calibrated adaptation framework for open-world MRI reconstruction grounded in the \textit{minimal intervention principle}: preserve transferable knowledge while calibrating only bias-prone components. Through transferability analysis, we observe that deeper layers and low-frequency components exhibit domain-specific characteristics while shallow layers and high-frequency components remain domain-agnostic, motivating frequency-guided prior calibration and regularization parameter adaptation. By intervening minimally and precisely, BiasRecon achieves state-of-the-art performance across cross-center, cross-anatomy, and cross-modality scenarios with fewer than 100 parameters.

%% file: bib.bib
@inproceedings{chungdecomposed,
  title={Decomposed Diffusion Sampler for Accelerating Large-Scale Inverse Problems},
  author={Chung, Hyungjin and Lee, Suhyeon and Ye, Jong Chul},
  booktitle={The Thirteenth International Conference on Learning Representations},
  year={2024}
}

@article{qu2012undersampled,
  title={Undersampled MRI reconstruction with patch-based directional wavelets},
  author={Qu, Xiaobo and Guo, Di and Ning, Bende and Hou, Yingkun and Lin, Yulan and Cai, Shuhui and Chen, Zhong},
  journal={Magnetic resonance imaging},
  volume={30},
  number={7},
  pages={964--977},
  year={2012},
  publisher={Elsevier}
}

@article{block2007undersampled,
  title={Undersampled radial MRI with multiple coils. Iterative image reconstruction using a total variation constraint},
  author={Block, Kai Tobias and Uecker, Martin and Frahm, Jens},
  journal={Magnetic Resonance in Medicine: An Official Journal of the International Society for Magnetic Resonance in Medicine},
  volume={57},
  number={6},
  pages={1086--1098},
  year={2007},
  publisher={Wiley Online Library}
}

@article{yang2018admm,
  title={ADMM-CSNet: A deep learning approach for image compressive sensing},
  author={Yang, Yan and Sun, Jian and Li, Huibin and Xu, Zongben},
  journal={IEEE transactions on pattern analysis and machine intelligence},
  volume={42},
  number={3},
  pages={521--538},
  year={2018},
  publisher={IEEE}
}

@article{aggarwal2018modl,
  title={MoDL: Model-based deep learning architecture for inverse problems},
  author={Aggarwal, Hemant K and Mani, Merry P and Jacob, Mathews},
  journal={IEEE transactions on medical imaging},
  volume={38},
  number={2},
  pages={394--405},
  year={2018},
  publisher={IEEE}
}

@article{zbontar2018fastmri,
  title={fastMRI: An open dataset and benchmarks for accelerated MRI},
  author={Zbontar, Jure and Knoll, Florian and Sriram, Anuroop and Murrell, Tullie and Huang, Zhengnan and Muckley, Matthew J and Defazio, Aaron and Stern, Ruben and Johnson, Patricia and Bruno, Mary and others},
  journal={arXiv preprint arXiv:1811.08839},
  year={2018}
}

@article{yaman2021zero,
  title={Zero-shot self-supervised learning for MRI reconstruction},
  author={Yaman, Burhaneddin and Hosseini, Seyed Amir Hossein and Akcakaya, Mehmet},
  journal={International Conference on Learning Representations},
  year={2022}
}

@article{wang2021denoising,
  title={Denoising auto-encoding priors in undecimated wavelet domain for MR image reconstruction},
  author={Wang, Siyuan and Lv, Junjie and He, Zhuonan and Liang, Dong and Chen, Yang and Zhang, Minghui and Liu, Qiegen},
  journal={Neurocomputing},
  volume={437},
  pages={325--338},
  year={2021},
  publisher={Elsevier}
}

@article{chung2022score,
  title={Score-based diffusion models for accelerated MRI},
  author={Chung, Hyungjin and Ye, Jong Chul},
  journal={Medical image analysis},
  volume={80},
  pages={102479},
  year={2022},
  publisher={Elsevier}
}

@article{jalal2021robust,
  title={Robust compressed sensing mri with deep generative priors},
  author={Jalal, Ajil and Arvinte, Marius and Daras, Giannis and Price, Eric and Dimakis, Alexandros G and Tamir, Jon},
  journal={Advances in Neural Information Processing Systems},
  volume={34},
  pages={14938--14954},
  year={2021}
}

@article{zhussip2019extending,
  title={Extending stein's unbiased risk estimator to train deep denoisers with correlated pairs of noisy images},
  author={Zhussip, Magauiya and Soltanayev, Shakarim and Chun, Se Young},
  journal={Advances in neural information processing systems},
  volume={32},
  year={2019}
}

@article{ramani2008monte,
  title={Monte-Carlo SURE: A black-box optimization of regularization parameters for general denoising algorithms},
  author={Ramani, Sathish and Blu, Thierry and Unser, Michael},
  journal={IEEE Transactions on image processing},
  volume={17},
  number={9},
  pages={1540--1554},
  year={2008},
  publisher={IEEE}
}

@article{uecker2014espirit,
  title={ESPIRiT—an eigenvalue approach to autocalibrating parallel MRI: where SENSE meets GRAPPA},
  author={Uecker, Martin and Lai, Peng and Murphy, Mark J and Virtue, Patrick and Elad, Michael and Pauly, John M and Vasanawala, Shreyas S and Lustig, Michael},
  journal={Magnetic resonance in medicine},
  volume={71},
  number={3},
  pages={990--1001},
  year={2014},
  publisher={Wiley Online Library}
}

@inproceedings{ong2018mridata,
  title={Mridata. org: An open archive for sharing MRI raw data},
  author={Ong, Frank and Amin, Shahab and Vasanawala, Shreyas and Lustig, Michael},
  booktitle={Proc. Intl. Soc. Mag. Reson. Med},
  volume={26},
  year={2018}
}

@article{wang2024cmrxrecon,
  title={CMRxRecon: A publicly available k-space dataset and benchmark to advance deep learning for cardiac MRI},
  author={Wang, Chengyan and Lyu, Jun and Wang, Shuo and Qin, Chen and Guo, Kunyuan and Zhang, Xinyu and Yu, Xiaotong and Li, Yan and Wang, Fanwen and Jin, Jianhua and others},
  journal={Scientific Data},
  volume={11},
  number={1},
  pages={687},
  year={2024},
  publisher={Nature Publishing Group UK London}
}

@article{liu2020highly,
  title={Highly undersampled magnetic resonance imaging reconstruction using autoencoding priors},
  author={Liu, Qiegen and Yang, Qingxin and Cheng, Huitao and Wang, Shanshan and Zhang, Minghui and Liang, Dong},
  journal={Magnetic resonance in medicine},
  volume={83},
  number={1},
  pages={322--336},
  year={2020},
  publisher={Wiley Online Library}
}

@article{yang2016sparse,
  title={Sparse reconstruction techniques in magnetic resonance imaging: methods, applications, and challenges to clinical adoption},
  author={Yang, Alice C and Kretzler, Madison and Sudarski, Sonja and Gulani, Vikas and Seiberlich, Nicole},
  journal={Investigative radiology},
  volume={51},
  number={6},
  pages={349--364},
  year={2016},
  publisher={LWW}
}

@inproceedings{ronneberger2015u,
  title={U-net: Convolutional networks for biomedical image segmentation},
  author={Ronneberger, Olaf and Fischer, Philipp and Brox, Thomas},
  booktitle={Medical image computing and computer-assisted intervention--MICCAI 2015: 18th international conference, Munich, Germany, October 5-9, 2015, proceedings, part III 18},
  pages={234--241},
  year={2015},
  organization={Springer}
}

@inproceedings{zhang2018perceptual,
  title={The Unreasonable Effectiveness of Deep Features as a Perceptual Metric},
  author={Zhang, Richard and Isola, Phillip and Efros, Alexei A and Shechtman, Eli and Wang, Oliver},
  booktitle={CVPR},
  year={2018}
}

@inproceedings{ulyanov2018deep,
  title={Deep image prior},
  author={Ulyanov, Dmitry and Vedaldi, Andrea and Lempitsky, Victor},
  booktitle={Proceedings of the IEEE conference on computer vision and pattern recognition},
  pages={9446--9454},
  year={2018}
}

@article{darestani2021accelerated,
  title={Accelerated MRI with un-trained neural networks},
  author={Darestani, Mohammad Zalbagi and Heckel, Reinhard},
  journal={IEEE Transactions on Computational Imaging},
  volume={7},
  pages={724--733},
  year={2021},
  publisher={IEEE}
}

@article{song2020denoising,
  title={Denoising diffusion implicit models},
  author={Song, Jiaming and Meng, Chenlin and Ermon, Stefano},
  journal={International Conference on Learning Representations},
  year={2021}
}

@article{knoll2019assessment,
  title={Assessment of the generalization of learned image reconstruction and the potential for transfer learning},
  author={Knoll, Florian and Hammernik, Kerstin and Kobler, Erich and Pock, Thomas and Recht, Michael P and Sodickson, Daniel K},
  journal={Magnetic resonance in medicine},
  volume={81},
  number={1},
  pages={116--128},
  year={2019},
  publisher={Wiley Online Library}
}

@article{geethanath2013compressed,
  title={Compressed sensing MRI: a review},
  author={Geethanath, Sairam and Reddy, Rashmi and Konar, Amaresha Shridhar and Imam, Shaikh and Sundaresan, Rajagopalan and DR, Ramesh Babu and Venkatesan, Ramesh},
  journal={Critical Reviews in Biomedical Engineering},
  volume={41},
  number={3},
  year={2013},
  publisher={Begel House Inc.}
}
